\documentclass{aa}

\usepackage[utf8]{inputenc}
\usepackage{graphicx}
\usepackage{amssymb}
\usepackage{amsmath}
\usepackage{float}
\usepackage{natbib}
\bibpunct{(}{)}{;}{a}{}{,} 
\usepackage{array}
\usepackage{hyperref}
\usepackage{subfig}
\usepackage[varg]{txfonts}
\usepackage{multirow}
\usepackage{siunitx}
\usepackage{upgreek}
\usepackage{textgreek}
\usepackage{ulem}
\usepackage{titlesec}

\usepackage{xcolor}

\hypersetup{colorlinks=true,citecolor=blue}

\titleformat{\paragraph}
{\normalfont\normalsize\bfseries}{\theparagraph}{1em}{}
\titlespacing*{\paragraph}
{0pt}{3.25ex plus 1ex minus .2ex}{1.5ex plus .2ex}

\def\app#1#2{%
  \mathrel{%
    \setbox0=\hbox{$#1\sim$}%
    \setbox2=\hbox{%
      \rlap{\hbox{$#1\propto$}}%
      \lower1.1\ht0\box0%
    }%
    \raise0.25\ht2\box2%
  }%
}

\DeclareTextSymbol{\degre}{T1}{6}
\DeclareTextSymbol{\degre}{OT1}{23}

\defcitealias{}{}          % X et al. 20XX

\begin{document}
  \title{High-resolution confirmation of an extended helium atmosphere around WASP-107b}
   \subtitle{}

   \author{   R. Allart     		  \inst{1,*}, 
  					V. Bourrier   	  \inst{1},  
   					C. Lovis     		  \inst{1}, 
   					D. Ehrenreich     \inst{1}, 
   					J. Aceituno     	  \inst{2,3}, 
   					A. Guijarro    	  \inst{2}, 
   					F. Pepe     		  \inst{1}, 
  					D. K. Sing   		  \inst{4,5}, 
   					J.J. Spake     		  \inst{4,5,6}, 
   					A. Wyttenbach	  \inst{7} 
               }
   
   \institute{\inst{1} Observatoire astronomique de l'Universit\'e de Gen\`eve, Universit\'e de Gen\`eve, 51 chemin des Maillettes, CH-1290 Versoix, Switzerland\\
   					\inst{2} Centro Astronomico Hispano Aleman, Sierra de los filabres sn, Gérgal, Almería, Spain.\\
   					\inst{3} Instituto de astrofísica de Andalucía (CSIC), Glorieta de la Astronomia sn, Granada Spain.\\
   					\inst{4} Department of Earth \& Planetary Sciences, Johns Hopkins University, Baltimore, MD, USA.\\
   					\inst{5} Department of Physics \& Astronomy, Johns Hopkins University, Baltimore, MD, USA.\\
   					\inst{6} Astrophysics Group, School of Physics, University of Exeter, Stocker Road, Exeter, EX4 4QL, UK.\\
   					\inst{7} Leiden Observatory, Leiden University, Postbus 9513, 2300 RA Leiden, The Netherlands.\\
              		* \email{romain.allart@unige.ch}
             }

   \date{Received December 19, 2018; accepted January 22, 2019}

  \abstract
   {Probing the evaporation of exoplanet atmospheres is key to understanding the formation and evolution of exoplanetary systems. The main tracer of evaporation in the UV is the Lyman-$\alpha$ transition, which can reveal extended exospheres of neutral hydrogen. Recently, the near-infrared (NIR) metastable helium triplet (10833 \AA) revealed extended thermospheres in several exoplanets, opening a new window into evaporation.}
   {We aim at spectrally resolving the first helium absorption signature detected in the warm Saturn WASP-107b with \textit{HST}/WFC3.}
   {We obtained one transit of WASP-107b with the high-resolution spectrograph CARMENES on the 3.5m telescope in Calar Alto.}
   {We detect an excess helium absorption signature of ~5.54$\pm$0.27 \% (20$\sigma$) in the planet rest frame during the transit. The detection is in agreement with the previous detection done with \textit{HST}/WFC3. The signature shows an excess absorption in the blue part of the lines suggesting that \ion{He}{i} atoms are escaping from the atmosphere of WASP-107b. We interpret the time-series absorption spectra using the 3D EVE code. Our observations can be explained by combining an extended thermosphere filling half the Roche lobe and a large exospheric tail sustained by an escape rate of metastable helium on the order of 10$^{6}$\,g$\cdot$s$^{-1}$. In this scenario, however, the upper atmosphere needs to be subjected to a reduced photoionisation and radiation pressure from the star for the model to match the observations.}
   {We confirm the presence of helium in the atmosphere of WASP-107b at high-confidence. The helium feature is detected from space and the ground. The ground-based high-resolution signal brings detailed information about the spatial and dynamical structure of the upper atmosphere, and simulations suggest that the \ion{He}{i} signature of WASP-107b probes both its thermosphere and exosphere establishing this signature as a robust probe of exoplanetary upper atmospheres. Surveys with NIR high-resolution spectrographs (e.g. CARMENES, SPIRou or NIRPS) will deliver a statistical understanding of exoplanet thermospheres and exospheres via the helium triplet.}

   \titlerunning{Confirmed helium atmosphere around WASP-107b}
   \authorrunning{R. Allart, V. Bourrier, C. Lovis et al.}
   \maketitle
%
%________________________________________________________________
% == Introduction ==
\section{Introduction}
Over the last decade, the exoplanetology field entered a phase of detailed characterisation of exoplanet atmospheres. The easiest targets for atmospheric studies are hot planets close to their host stars, like hot Jupiters and warm Neptunes, whose hydrogen- and helium-dominated atmospheres can reach large scale heights. Atmospheres heated through absorption of the stellar irradiation can expand hydrodynamically \citep{vidal-madjar_extended_2003,lammer_atmospheric_2003}. Hydrogen, through the Lyman-$\alpha$ line at ultra-violet (UV) wavelengths, was first observed escaping from hot Jupiters such as HD209458b \citep{vidal-madjar_extended_2003} and HD189733b \citep{lecavelier_des_etangs_evaporation_2010} and forming extended cometary tails. More recently, even larger hydrogen exospheres were detected around the warm-Neptunes GJ436b \citep{ehrenreich_giant_2015} and GJ3470b \citep{bourrier_hubble_2018}. \\

Warm-Neptunes are important targets for atmospheric studies because many of them stand at the edge of the evaporation desert \citep{lecavelier_des_etangs_diagram_2007,beauge_emerging_2013}, a lack of Neptune-mass planets at short orbital distance. This desert can be explained by planets that are not massive enough to retain their escaping gaseous atmosphere. Probing the gas escaping from planets around the desert is thus particularly important to understand the evolution of the close-in planet population. While the Lyman-$\alpha$ line is an excellent tracer of the outermost atmospheric layers, it suffers from the lack of stellar continuum in the UV and from interstellar medium (ISM) absorption, limiting observations to extrasolar systems close to the Sun. \\

The metastable helium triplet, located in a spectral region devoid of strong ISM absorption (\citet{indriolo_interstellar_2009}, 10832.1, 10833.2 and 10833.3 \AA\ in the vacuum)  with a bright continuum, was theorized as a potential tracer of upper atmospheres \citep{seager_theoretical_2000,oklopcic_new_2018}, and was recently detected in several exoplanets: the warm-Saturn WASP-107b \citep{spake_helium_2018}, the hot Jupiters WASP-69b and HD189733b \citep{nortmann_ground-based_2018,salz_detection_2018} and the smaller warm Neptune HAT-P-11b \citep{allart_spectrally_2018,mansfield_detection_2018}. All these planets, except WASP-107b, have been observed at high resolution with CARMENES, allowing us to trace the presence of \ion{He}{i} down to the thermosphere. \\

WASP-107b (\cite{anderson_discoveries_2017}) is a warm Saturn (0.11\,$\pm$\,0.01\,$M_{J}$ and 0.94\,$\pm$\,0.02\,$R_{J}$) located at the upper-radius border of the evaporation desert. It orbits a K6-type star with a period of 5.72 days (physical and orbital parameter are given in Table \ref{parameter used}). It has twice the mass of Neptune but has a radius similar to Jupiter, and is thus one of the less dense exoplanets (0.19\,$\pm$\,0.03\,g$\cdot$cm$^{-3}$), well-suited for atmospheric characterisation. So far two species have been detected in its atmosphere, water \citep{kreidberg_water_2018} and  helium \citep{spake_helium_2018}, both with \textit{Hubble Space Telescope} (\textit{HST}/WFC3). The helium feature, however, was unresolved and strongly diluted because it was measured with the G102 grism, which has a low resolution of 67 \AA\ (the typical width of the absorption helium feature is about 1 \AA) and conservative 98 \AA\ bins have been used. Thus, the absorption profile of the \ion{He}{i} absorption signature from WASP-107b remains poorly characterised. Observations of the helium triplet in WASP-107b at high spectral resolution are thus required to resolve the lines and constrain the properties of the metastable helium population around the planet. Here, we analyse one transit of WASP-107b obtained with the high-resolution spectrograph CARMENES to spectrally resolve the helium triplet.

%________________________________________________________________

\section{CARMENES observations}

\subsection{Reduced data}
\label{CARMENES_obs}
We observed one transit of WASP-107b on the 23$^\mathrm{rd}$ of February 2018 (DDT.S18.188; PI: Allart) with the CARMENES high-resolution spectrograph \citep{quirrenbach_carmenes_2014} on the 3.5m telescope at Calar Alto. The transit dataset consists in 20 spectra with exposure time of 956\,s, with 10 spectra covering the 2.75 hours of the transit. The signal to noise ratio (SNR) varies from 26 to 52 within the Echelle order 55 (10804-11005 \AA). Two spectra obtained at the end of the night have been excluded from the following analysis due to a low SNR, below 30. Data were automatically reduced with the CARMENES Reduction and Calibration pipeline \citep{caballero_carmenes:_2016}, which applies a bias, flat and cosmic ray correction of the raw spectra, and then a flat-relative optimal extraction (FOX; \cite{zechmeister_flat-relative_2014}) and wavelength calibration \citep{bauer_calibrating_2015}. The resulting output is defined in the Earth laboratory frame and is composed of wavelength (defined in vacuum), flux and flux uncertainty maps (order vs. pixel number). Our analysis focuses on the order 55, which includes the helium triplet.

\subsection{Telluric correction}
\begin{figure}[h]
\resizebox{\hsize}{!}{\includegraphics{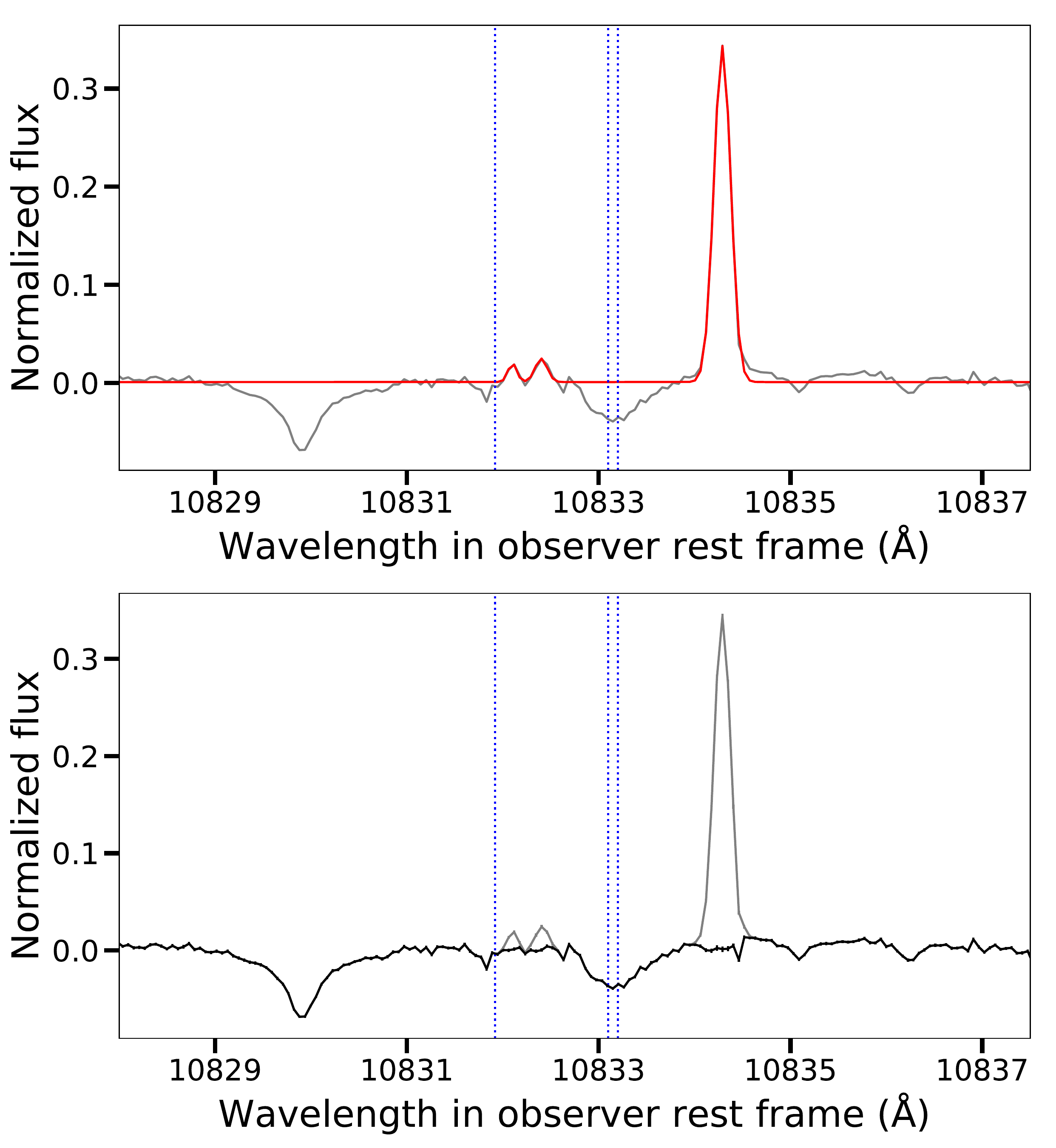}}
\caption[tellurique]{Master-out stellar spectrum in the observer frame (grey profile). Blue dotted lines indicate the \ion{He}{i} triplet.  \textit{Top panel:} The best-fit to the telluric OH emission lines is shown in red. \textit{Bottom panel:} Master-out spectrum corrected for the best-fit model of the OH lines (black profile).}
\label{tellurique}
\end{figure}

Ground-based spectra are contaminated by the Earth's atmosphere manifested as telluric absorption lines (including water lines) and as telluric emission lines (such as OH). Both set of lines are present near the helium triplet in our observations (Fig. \ref{tellurique}). \\
The closest water absorption line is at 10835.1 \AA\ in the observer rest frame and is redshifted by more than 2 \AA\ from the planet helium triplet (see Fig. \ref{tellurique}). Since this is outside of the spectral range required for our analysis, we did not correct for this telluric line.\\
OH emission lines are clearly visible in Fig. \ref{tellurique} at 10832.1, 10832.4 and 10834.3 \AA\ and fall close to the stellar metastable helium lines at 10831.9, 10833.1 and 10833.2 \AA\ (wavelengths are given in the observer rest frame). During the observation the second fiber of the spectrograph, fiber B, was put on the sky to monitor the emission lines. We built a high-SNR master-sky spectrum by co-adding all fiber B spectra. We confirmed the telluric origin of the OH lines, their wavelength positions, and we excluded the presence of other emission lines.\\
To correct for these emission lines, we built a master-out spectrum from out-of-transit spectra (phase < -0.01263) in the observer rest frame. We then fit a Voigt profile for the line at 10834.3 \AA\ and Gaussian profiles for the two shallowest lines (10832.1 and 10832.4 \AA). Finally, we fit the derived profiles to the emission lines in each individual exposure using a scaling factor. Best-fit models were then subtracted from each spectrum.

%________________________________________________________________

\section{Analysis of the observations}
We used a similar method as \cite{allart_spectrally_2018} to build the transmission spectrum, which consists in Doppler-shifting the telluric-corrected spectra from the observer to the stellar rest frame by accounting for the barycentric Earth radial velocity, the stellar reflex motion of the star and its systemic velocity (Table \ref{parameter used}). Then, spectra obtained out-of-transit are co-added to form a master-out spectrum, which is normalized to unity by reference bands in the blue (10821.9 to 10827.1 \AA) and the red (10838.4 to 10841.6 \AA) side of the helium triplet. Similarly all spectra have been normalised by the aforementioned reference bands for comparison. Two spectra (02h43 and 03h02 UT respectively at phase -0.0032 and -0.0009) exhibit emission features over 3 pixels from 10829.8 to 10830.3 \AA\ and one spectrum (03h53 UT, at phase 0.0053) has also an emission feature over 4 pixels from 10835.6 to 10836.0 \AA. Those features are likely due to cosmic hits. We replaced the flux value of these contaminated pixels by the median of the corresponding pixel over the night.\\

\begin{figure}[h]
\resizebox{\hsize}{!}{\includegraphics{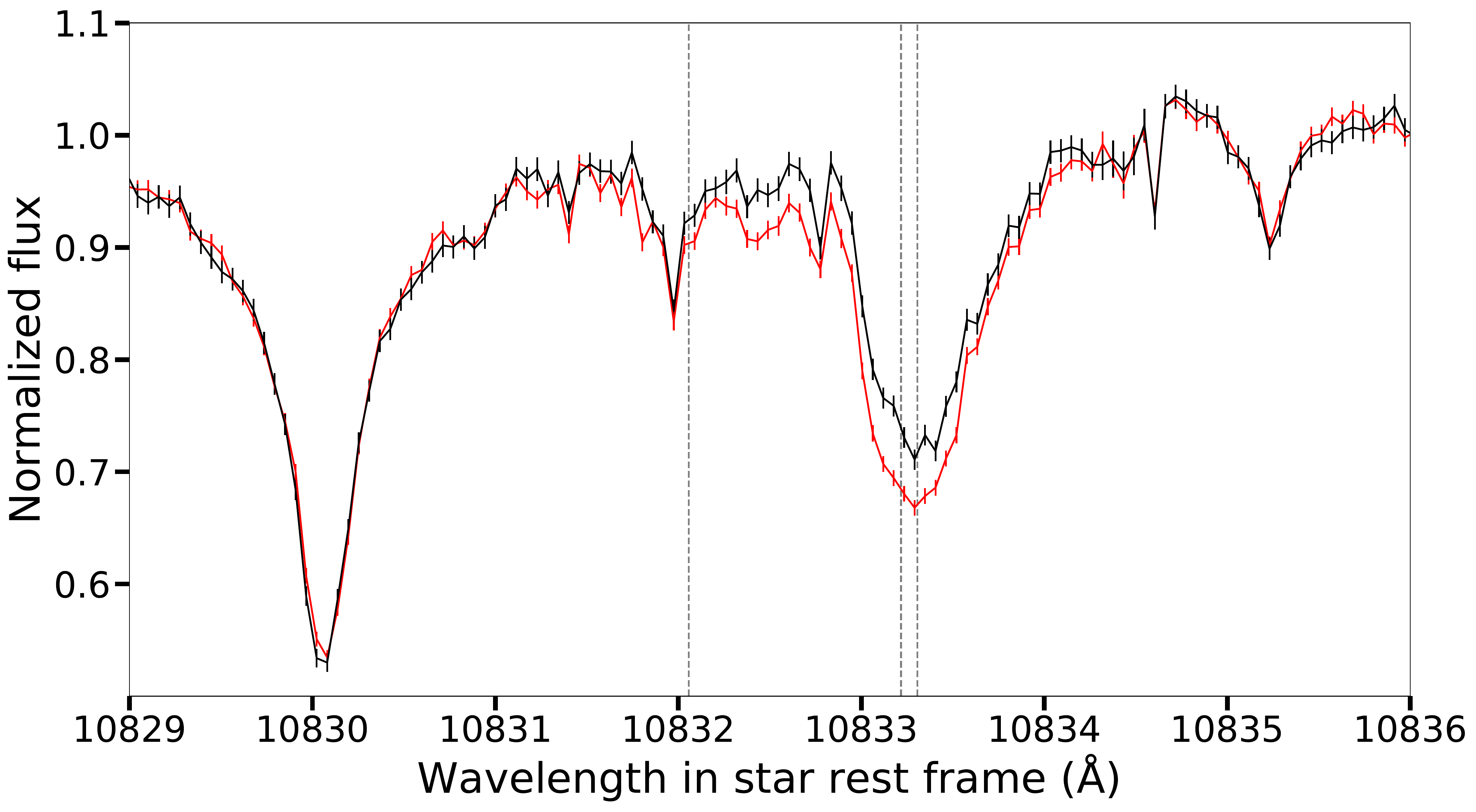}}
\caption[Master]{Master-out (black) and Master-in (red) spectra in the star frame. Vertical grey lines indicate the position of the helium triplet lines. In-transit absorption is clearly visible by eye.}
\label{master}
\end{figure}

We first performed a visual comparison between the Master-out spectrum and a Master-in spectrum obtained by co-adding all normalized spectra between t$_\mathrm{II}$ and t$_\mathrm{III}$ (Fig. \ref{master}). An absorption signature is clearly visible by eye in the stellar spectrum, at the location of the helium triplet.\\

\begin{figure}[h]
\resizebox{\hsize}{!}{\includegraphics{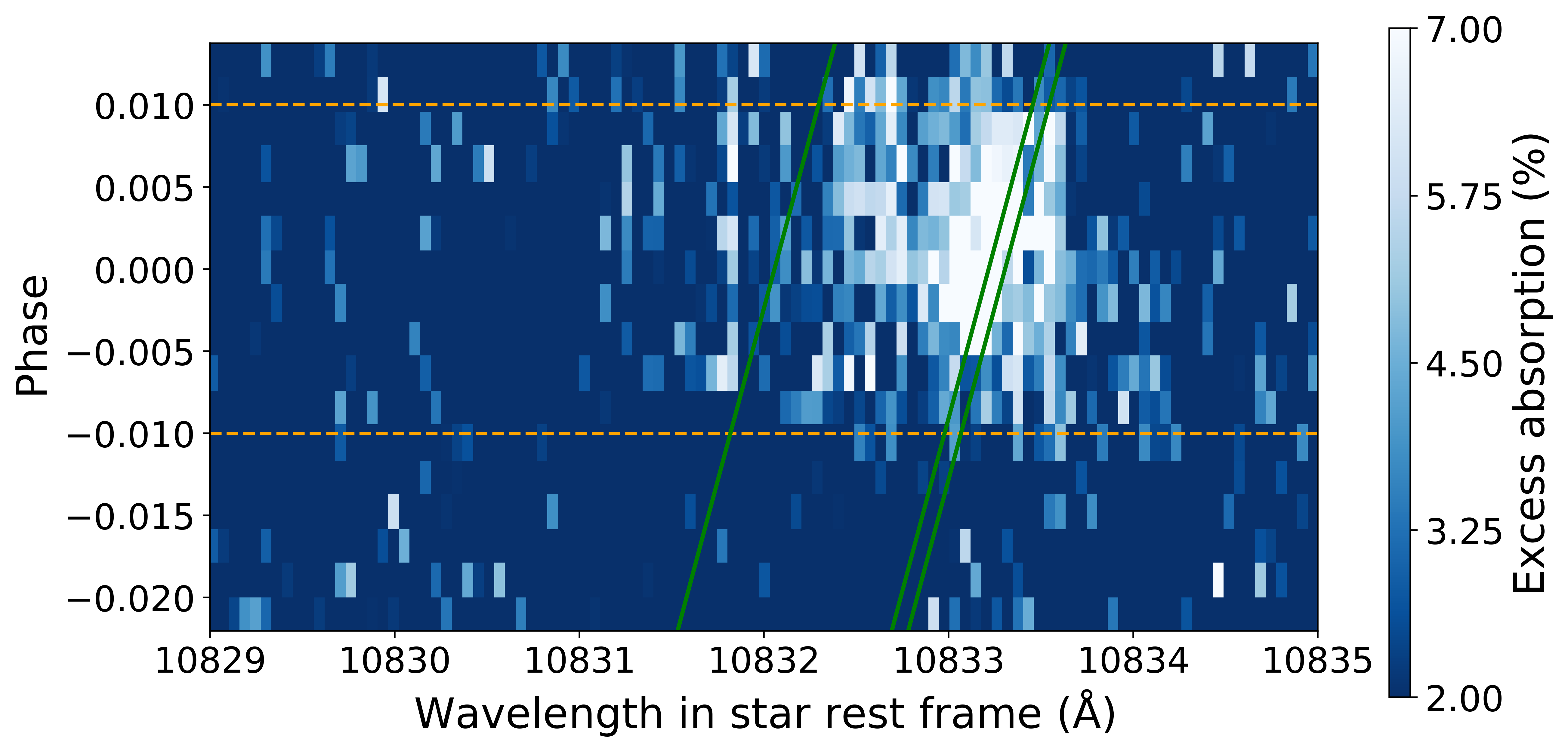}}
\caption[Map]{Phase vs wavelength map of ITS in the star frame. An excess absorption signature (in white) is visible during the transit (first and fourth contacts are highlighted in orange). Green curves are the helium planetary tracks.}
\label{map}
\end{figure}

To further investigate the origin of the signature, we divided all spectra by the master out and hereafter referred to them as the Individual Transmission Spectra (ITS). Fig. \ref{map} is a phase vs. wavelength map of the ITS in the stelllar rest frame. The excess absorption signature is visible during the transit only, and follows the radial velocity motion of the planet. This shows unambiguously the planetary origin of the helium absorption signature, confirming the unresolved detection from \cite{spake_helium_2018}. 

\begin{figure}[h]
\resizebox{\hsize}{!}{\includegraphics{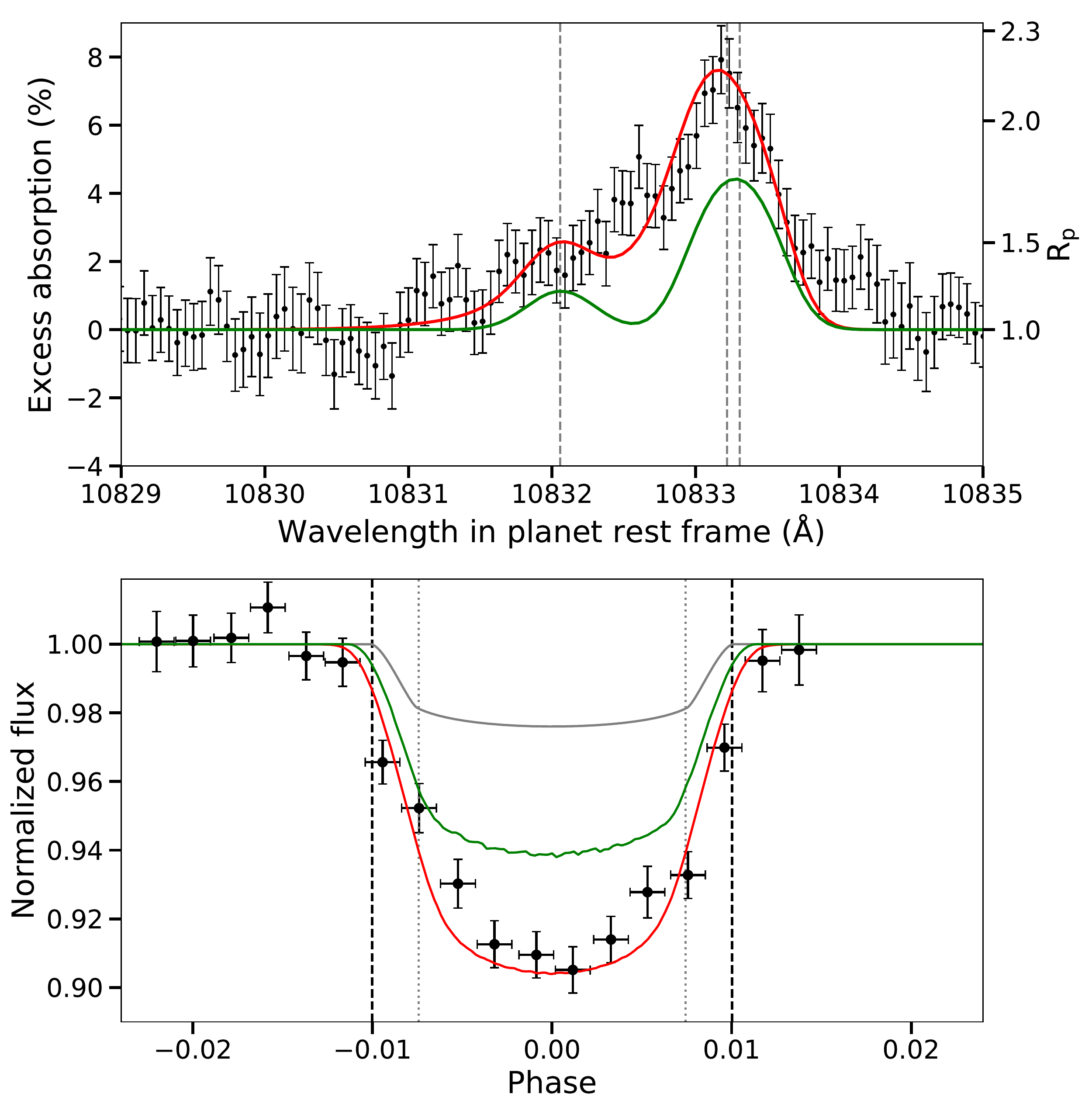}}
\caption[TS_&_lc]{\textit{Top panel:} Transmission spectrum of WASP-107b in the region of the \ion{He}{i} triplet (black points), in the planet frame. The red line shows the theoretical profile obtained with the EVE code for the model shown in Fig \ref{fig:model_views}. The green line shows the contribution from the model thermosphere alone. The three vertical grey dashed lines correspond to the helium triplet transition. \textit{Bottom panel:} Helium light curve integrated from 10832.80 to 10833.55 \AA\, from the observations (black), the theoretical atmospheric continuum (grey) obtained with Batman, and the simulated EVE atmosphere (same color code as the top panel). The two vertical black dashed lines correspond to the contact points t$_\mathrm{I}$ and t$_\mathrm{IV}$ while the two vertical grey dotted lines correspond to t$_\mathrm{II}$ and t$_\mathrm{III}$.}
\label{TS&lc}
\end{figure}

 The ITS were then scaled by the reference light curve and shifted to the planet frame. We used the Batman package \citep{kreidberg_batman:_2015} to build a reference light curve based on the transit depth and non-linear limb darkening coefficients derived by \cite{spake_helium_2018} from \textit{HST}/WFC3 observations in the continuum around the helium triplet, and orbital properties derived by \cite{anderson_discoveries_2017,dai_oblique_2017} (Table \ref{appendiceA}).  To characterise the spectro-temporal properties of the helium absorption we built a helium light curve, i.e. the temporal variation of the flux in the spectral range absorbed by the planetary atmosphere, by integrating them from 10832.80 to 10833.55 \AA. We also built a master transmission spectrum as the average of the seven ITS obtained between t$_\mathrm{II}$ and t$_\mathrm{III}$, selected to maximise the significance of the helium feature. The absorption depth from the atmospheric continuum was removed from this transmission spectrum (Fig. \ref{TS&lc}, top panel) to obtain the excess absorption spectrum by helium in the atmosphere. \\

Fig. \ref{TS&lc} shows the transmission spectrum and the helium light curve. There is a significant helium excess absorption of 5.54 $\pm$ 0.27 \% (20-$\sigma$ detection) over a 0.75 \AA\ passband centered  around the peak of excess absorption, which reaches 7.92 $\pm$ 1.00 \% (or an equivalent opaque radius of 2.2 R$_p$) at 10833.1768 \AA. The helium signature shows a strong spectral asymmetry, with excess absorption in the blue part of the lines. The helium light curve is roughly symmetrical in time and centered around mid-transit, but as the effective planet radius in the \ion{He}{i} lines is larger than the atmospheric continuum, the transit duration is longer by about 30 minutes.  This further proves the \ion{He}{i} lines of planetary origin, as a false positive signature arising from stellar activity does not change the transit duration.\\

The star WASP-107 is a slow rotator ($ v_{eq} \cdot sin(i)$=2.5 km$\cdot$s$^{-1}$ , \citet{anderson_discoveries_2017}), and the Rossiter-McLaughlin (RM) of WASP-107b was not detected in velocimetry. The impact of the RM effect in the transmission spectrum is thus expected to be limited. We confirmed that it was below the 10$^{-3}$ level by performing two simulations with the EVE code (see section \ref{discussion}) assuming an obliquity of $\lambda$= 40\,$^{\circ}$ or $\lambda$=140\,$^{\circ}$ (the range within which the obliquity of WASP-107b is currently known, \cite{dai_oblique_2017}).\\

\begin{figure}[h]
\resizebox{\hsize}{!}{\includegraphics{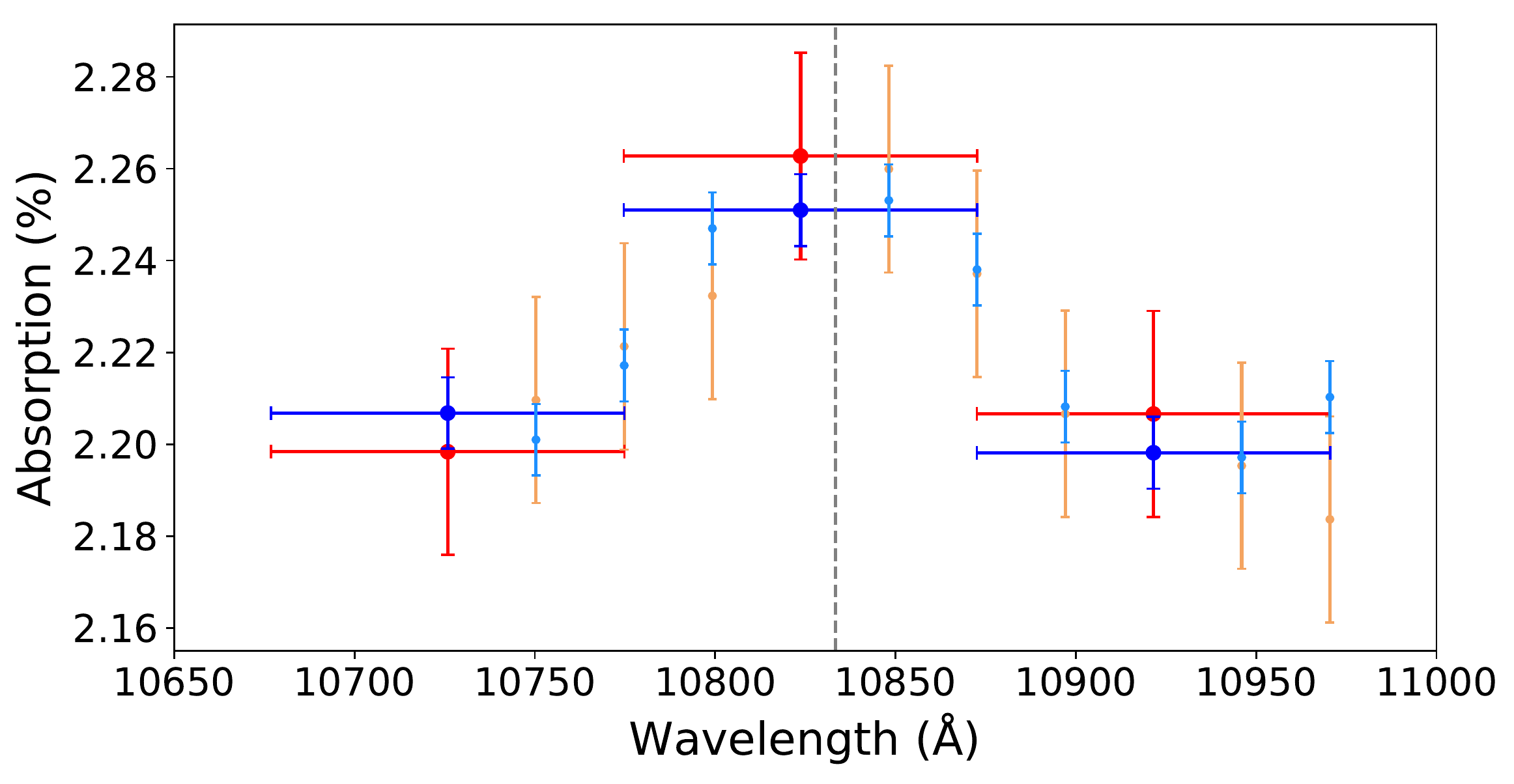}}
\caption[HST]{Comparison between the \textit{HST}/WFC3 dataset of \cite{spake_helium_2018} in blue with our degraded high-resolution CARMENES dataset in red. Transmission spectra are binned over 98 \AA. Lighter points correspond to the same transmisssion spectra but shifted by 24.5, 49.0 and 73.5 \AA. The vertical grey dashed line is the helium triplet transition.}
\label{HST}
\end{figure}

To compare our results with the HST low-resolution transmission spectrum of WASP-107b obtained by \cite{spake_helium_2018}, we convolved the CARMENES high-resolution transmission spectrum to the resolution of WFC3 and binned it over 98 \AA\ pixels. Both transmission spectra have been created using exposures obtained between phases -0.0063 to -0.0008. The transmission spectra are compatible within the 1-$\sigma$ errors (Fig. \ref{HST}) and thus the signal is repeatable over two different epochs. WASP-107b is one of the few planets to have the same species detected in its upper atmosphere with both ground-based and space-borne instruments.\\

% == Discussion ==

\section{Interpretation of the \ion{He}{i} signature}
\label{discussion}

Because it was not spectrally resolved, the signature of helium detected in the HST data of WASP-107b could be well fitted with two very different models for its upper atmosphere \citep{spake_helium_2018}. A 1D model of a hot thermosphere extending far beyond the Roche Lobe (based on \cite{oklopcic_new_2018}) yielded deep absorption lines, symmetrical in the planet rest frame. In contrast, a 3D model of an exospheric tail (based on \cite{bourrier_3d_2013}) yielded shallow absorption lines, spread over a wide spectral range because of the dynamics of helium atoms blown away from the planet. Our time-series of high-resolution CARMENES spectra allows us to study in details the profile of the absorption signature and its evolution over the transit (Fig.~\ref{map}). As aforementioned, the \ion{He}{i} absorption lines are asymmetrical, with a clear excess absorption at blueshifted velocities in the planet rest frame (Fig.~\ref{TS&lc}). This suggests that the outer atmospheric layers are being blown away from WASP-107b, forming a comet-like tail that seems at odds with the symmetry of the helium light curve (Fig.~\ref{TS&lc}). In fact this was expected from the 3D simulations performed in \cite{spake_helium_2018}, which showed that radiation pressure on escaping metastable helium atoms is so strong that a tail is formed which is aligned with the star-planet axis, with a roughly circular projection in the plane of sky. However, the measured absorption extends over a shorter wavelength range than predicted by the tail simulations from \cite{spake_helium_2018}, and shows a deep core centered on the line transitions in the planet rest frame (Fig.~\ref{TS&lc}). This similarity with the theoretical profile from the 1D model in \cite{spake_helium_2018} suggests that part of the signal arises from helium in an extended thermosphere surrounding WASP-107b. Therefore, our observations might probe for the first time both the thermosphere and the exosphere of an exoplanet. \\

To further assess this possibility, we used the version of the EVaporating Exoplanet code (EVE) \citep{bourrier_3d_2013,bourrier_evaporating_2016} presented in \cite{spake_helium_2018,allart_spectrally_2018}. The planetary system is simulated in 3D in the star rest frame, and the code calculates theoretical spectra comparable to the CARMENES observations during the transit of the planet and its atmosphere. The thermosphere is modeled as a parameterized grid, using density and velocity profiles calculated with a spherically symmetric, steady-state isothermal wind model \citep{parker_dynamics_1958,oklopcic_new_2018}. The exosphere is modeled by releasing metastable helium atoms at the top of the thermosphere, and computing their dynamics with Monte-Carlo particle simulations accounting for the planet and star gravity, and the stellar radiation pressure. Metastable helium atoms in the simulation can be photoionized by the stellar incident radiation, or radiatively de-excited into their fundamental state. The density profile of metastable helium in the thermosphere is scaled so that it matches the density of exospheric metaparticles at the exobase. \\

\begin{figure}     %gauche    bas   droite
\includegraphics[trim=0cm 0cm 0cm 0cm,clip=true,width=\columnwidth]{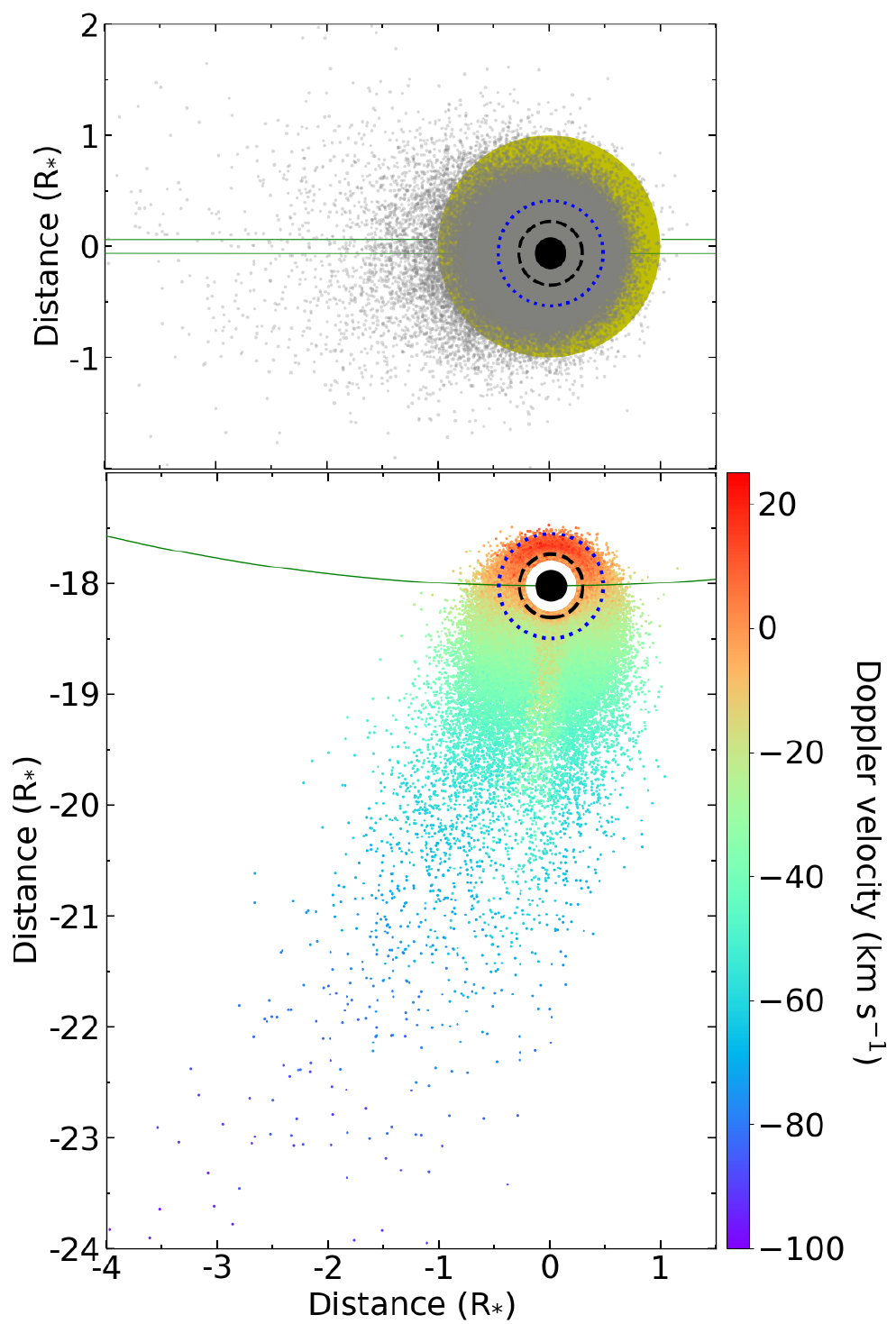}
\caption[]{3D model of WASP-107b upper atmosphere simulated with the EVE code. Particles of metastable helium escape at the exobase (shown in projection as a dashed black circle), which separates the simulated thermosphere and exosphere. The planet as seen in the near-infrared continuum is shown as a black disk. The dotted blue line shows the projection of the Roche Lobe. The green curve shows the  orbit of WASP-107b. \textit{Top panel}: view along the line-of-sight toward Earth, with metastable helium particles shown in grey. The yellow disk is the star. \textit{Bottom panel}: view from the above of the planetary system. Metastable helium particles are colored as a function of their radial velocity, and only shown within the orbital plane for the sake of clarity.}
\label{fig:model_views}
\end{figure}

Knowledge of the stellar flux is required below 2593\,\AA\, to calculate the photoionization rate of metastable helium atoms, and in the region of the \ion{He}{i} triplet to calculate radiation pressure \citep{spake_helium_2018,allart_spectrally_2018}. In a first approach we used the semi-synthetic spectrum of the K6-type star HD85512 as a proxy for WASP-107, replacing it in the region of the \ion{He}{i} triplet by the out-of-transit CARMENES spectrum rescaled to the flux level expected for a blackbody at the temperature of WASP-107. However, escaping atoms in the simulations were then accelerated so fast that they yielded excess absorption over a much larger velocity range than observed in the data. \\

We found that decreasing artificially the stellar spectrum in the region of the \ion{He}{i} triplet by a factor 50, so that radiation pressure is about 50\% stronger than the star gravity, leads to the formation of an exospheric tail with a velocity gradient consistent with the data (Fig.~\ref{TS&lc}). In this scenario, however, escaping helium atoms are subjected to a lower acceleration and are photo-ionized before they can reach the observed velocities. We were able to retrieve the observed shape of the absorption profile to a reasonable degree by decreasing the XUV flux by the same factor as the near-infrared flux (Fig.~\ref{TS&lc}). In that case the lifetime of metastable helium atoms becomes controlled by radiative de-excitation ($\sim$131 min) rather than photo-ionization ($\sim$7\,min with the original XUV flux at the semi-major axis of 18.02\,$\pm$\,0.27\,$R_{*}$). It is possible that the XUV spectrum of HD\,85512 is not a good proxy for WASP-107, especially since the EUV portion is derived empirically \citep{france_muscles_2016}. However we see no reason for the intrinsic stellar spectrum of WASP-107 to be lower than the expected black body in the region of the \ion{He}{i} triplet. Our simulations thus suggest that additional physical processes shape the population of metastable helium atoms in the upper atmosphere of WASP-107b. For example they could be shielded from incoming photons by an unkown absorber, or their dynamics could be affected by collisions with other escaping species or with stellar wind particles. \\

Assuming that simulations with a reduced photo-ionization and radiation pressure capture the overall structure of WASP-107b exosphere, it is interesting to look at the upper atmospheric properties required to explain the data. The model in Fig.~\ref{fig:model_views} was obtained by setting the exobase at 2\,R$_{p}$, so that both the thermosphere and the exosphere contribute to the theoretical absorption profile (Fig.~\ref{TS&lc}). With a thermospheric temperature of 12000\,K (assuming a solar-like hydrogen-helium composition with a mean atomic weigth of 1.2), metastable helium atoms escape at the exobase with a thermal wind velocity of $\sim$12\,km$\cdot$s$^{-1}$. They fill the Roche Lobe (3.3\,Rp) before being blown away by radiation pressure (Fig.~\ref{fig:model_views}). The absorption profile was well reproduced with an escape rate of metastable helium of about 8$\times$10$^{5}$\,g$\cdot$s$^{-1}$, which is in between the mass loss rates derived from the HST data by the 1D thermospheric and 3D exospheric models in \cite{spake_helium_2018}.\\

%________________________________________________________________

% == Conclusion ==
\section{Conclusion}
\label{conclusion}

\cite{spake_helium_2018} have detected absorption in the \ion{He}{i} triplet with HST during a single transit of the warm Neptune WASP-107b. Due to the low spectral resolution of WFC3 the absorption line profile could not be resolved, and the exact origin of helium in the planet atmosphere remained unclear. The strong dilution of helium absorption signatures measured with \textit{HST}/WFC3 limits its detection with this instrument to exoplanets with the deepest atmospheric transits like WASP-107b. Using ground-based CARMENES observations, we detect and resolve the signature of helium at high confidence (5.54 $\pm$ 0.27 \%, 20-$\sigma$) during a single transit of WASP-107b. The absorption signature is located at the \ion{He}{i} triplet wavelengths in the planet rest frame, displays a longer transit duration than the atmospheric continuum, and occurs during the planetary transit, showing unambiguously its planetary origin and confirming the result of \cite{spake_helium_2018}. \\

WASP-107b is the second exoplanets for which helium has been detected from the ground and from space after HAT-P-11b \citep{allart_spectrally_2018,mansfield_detection_2018}. While the helium absorption shows no temporal asymetry, its signature in the planet rest frame is asymetric with excess absorption in the blue parts of the lines. Using 3D numerical simulations with the EVE code, we explain the observations by extended thermosphere sustaining an exospheric comet-like tail. This scenario requires that escaping helium atoms are blown away by a reduced radiation pressure, suggesting that additional physical mechanisms are at play. \\

Further observations at high spectral resolution are required to get more insight into the dynamics and structure of WASP-107b upper atmosphere but also to investigate possible temporal variability of the signature (see the case of HAT-P-11, \cite{allart_spectrally_2018}). Our observations nonetheless show clearly that the near-infrared helium triplet can trace both the thermosphere and the exosphere, in contrast to observations of other exoplanets in this line \citep{allart_spectrally_2018,nortmann_ground-based_2018,salz_detection_2018}. Combined with the new generations of NIR high-resolution spectrographs (e.g. CARMENES, SPIRou, NIRPS) and their large atmospheric surveys, the \ion{He}{i} triplet will usher a new era of statistical studies of exoplanet extended atmospheres.

%________________________________________________________________

% == Acknowledgements ===
\begin{acknowledgements}
We thank the anonymous referee for the careful reading and comments. We acknowledge the Geneva exoplanet atmosphere group for fruitful discussions. This work has been carried out within the frame of the National Centre for Competence in Research 'PlanetS' supported by the Swiss National Science Foundation (SNSF). R.A., V.B., C.L., D.E., F.P. acknowledge the financial support of the SNSF. This project has received funding from the European Research Council (ERC) under the European Union's Horizon 2020 research and innovation programme (project FOUR ACES; grant agreement No 724427). A.W. acknowledge the financial support of the SNSF by the grant number P2GEP2\_178191. This work was based on observations collected at the Centro Astronómico Hispano Aleman (CAHA), operated jointly by the Max-Planck Institut fur Astronomie and the Instituto de Astrofisica de Andalucia (CSIC) under DDT proposal. 
\end{acknowledgements}
%-------------------------------------------------------------------
% == References ==
\bibliographystyle{aa}
\bibliography{bib}

\begin{appendix}
\section{Appendix A}
\label{appendiceA}

We refined the planetary mass, the radial velocity semi-amplitude and the semi-major axis of the orbit using the DACE platform \citep{delisle_analytical_2016,diaz_harps_2016} based on the existing radial velocity data points obtained with Coralie and HARPS. We used a Metropolis-Hastings Markov chain Monte Carlo algorithm with Gaussian priors based on previous studies \citep{anderson_discoveries_2017,dai_oblique_2017}.

\begin{table}[h]
\centering
\caption{\footnotesize Adopted physical and orbital parameters of WASP-107b.}\label{parameter used}
\begin{tabular}{lccr}
\hline
Parameter & Symbol & Value & Reference \\
\hline
Stellar radius & $R_{*}$ & 0.66\,$\pm$\,0.02\,$R_{\odot}$ & \cite{anderson_discoveries_2017} \\

Planet radius & $R_{p}$ &  0.94\,$\pm$\,0.02\,$R_{J}$ & \cite{anderson_discoveries_2017} \\

White-light radius ratio & $R_{p}/R_{*}$ & 0.142988\,$\pm$0.00012\, & \cite{spake_helium_2018} \\

Stellar mass & $M_{*}$ & 0.69\,$\pm$\,0.05\,$M_{\odot}$ & \cite{anderson_discoveries_2017} \\

Planet mass & $M_{p}$ & 0.11\,$\pm$\,0.01\,$M_{J}$ & This work, DACE \\

Epoch of transit & $T_{0}$ & 2457584.329897\,$\pm$\,0.000032\,BJD & \cite{dai_oblique_2017} \\

Duration of transit & $T_{14}$ & 0.1147\,$\pm$\,0.0003\,d  & \cite{anderson_discoveries_2017} \\

Orbital period & $P$ & 5.721474\,$\pm$\,0.000004\,d & \cite{dai_oblique_2017}\\

Systemic velocity & $\gamma$ & 14137.88\,$\pm$\,1.80 \,m\,s$^{-1}$ & This work, DACE \\

Semi-amplitude & K$_*$ & 16.45\,$\pm$\,1.21\,m\,s$^{-1}$ & This work, DACE \\

Eccentricity & $e$ & 0.0 & Fixed \\

Argument of the periastron &$\omega$ & 0.0 & Fixed \\

Semi-major axis & $a$ & 18.02\,$\pm$\,0.27\,$R_{*}$ & This work, DACE \\

Inclination & $i$ & 89.8\,$\pm$\,0.2\,$^{\circ}$ & \cite{dai_oblique_2017}\\

Limb-darkening & $u_1$ & 0.72615373 & \cite{spake_helium_2018} \\

Limb-darkening  & $u_2$ &  -0.70859436 & \cite{spake_helium_2018} \\

Limb-darkening  & $u_3$ & 1.09027178 & \cite{spake_helium_2018} \\

Limb-darkening  & $u_4$ &  -0.45320191 & \cite{spake_helium_2018} \\

\hline
\end{tabular}
\end{table}

\end{appendix}
\end{document}